\documentclass{aa}  

\usepackage{graphicx}
\usepackage{txfonts}
\usepackage{xcolor} 
\usepackage{amssymb}
\usepackage{soul} 
\usepackage[normalem]{ulem}
\usepackage{float}
\usepackage{ragged2e}
\usepackage{caption}
\usepackage{subcaption}
\usepackage{rotating}
\usepackage{psfrag}
\usepackage{graphicx}
\usepackage{txfonts}

\usepackage{textcomp}
\begin{document} 

   \titlerunning{Target selection of VVV Bulge galaxies}
   \authorrunning{Duplancic et al.}
   \title{Enlightening the Universe behind the Milky Way Bulge}

   \subtitle{I. Target selection of VVV Bulge galaxies}

   \author{Fernanda Duplancic\inst{1}\thanks{fduplancic@unsj-cuim.edu.ar}, 
          Sol Alonso\inst{1},
          Georgina Coldwell\inst{1},
          Daniela Galdeano\inst{1},
          Dante Minniti\inst{2,3},
          Julia Fernandez\inst{1},
          Valeria Mesa\inst{4,5,6},
          Noelia Perez\inst{1},
          Luis Pereyra\inst{7,8}
          \and
          Franco Pavesich\inst{9}
          }

  \institute{Departamento de Geof\'{i}sica y Astronom\'{i}a, CONICET, Facultad de Ciencias Exactas, F\'{i}sicas y Naturales, Universidad Nacional de San Juan, Av. Ignacio de la Roza 590 (O), J5402DCS, Rivadavia, San Juan, Argentina.       
    \and
    Instituto de Astrof\'isica, Facultad de Ciencias Exactas, Universidad Andres Bello, Av. Fernandez Concha 700, Las Condes, Santiago, Chile.
    \and
    Vatican Observatory, V00120 Vatican City State, Italy.
    \and
    Instituto de Investigación Multidisciplinar en Ciencia y Tecnología, Universidad de La Serena, Raúl Bitrán 1305, La Serena, Chile.
    \and
    Association of Universities for Research in Astronomy (AURA), Av. Juan Cisternas 1500,  La Serena, Chile
    \and
    Grupo de Astrofísica Extragaláctica-IANIGLA, CONICET, Universidad Nacional de Cuyo (UNCuyo), Gobierno de Mendoza. Parque Gral San Martín, CC 330, CP 5500, Mendoza, Argentina.
    \and
    Instituto de Astronomía Teórica y Experimental (IATE-CONICET), Laprida 854, X5000BGR, C\'ordoba, Argentina.
    \and
    Observatorio Astron\'omico de C\'ordoba, Universidad Nacional de C\'ordoba (OAC-UNC), Laprida 854, X5000BGR, C\'ordoba, Argentina.
    \and
    Departamento de Geof\'{i}sica y Astronom\'{i}a, Facultad de Ciencias Exactas, F\'{i}sicas y Naturales, Universidad Nacional de San Juan, Av. Ignacio de la Roza 590 (O), J5402DCS, Rivadavia, San Juan, Argentina. 
    }
   \date{Received xxx; accepted xxx}

 
  \abstract
   {The location of the Solar System constrains the detection of extragalactic sources beyond the Milky Way plane. The optical observations are hampered in the so--called Zone of Avoidance (ZOA) where stellar crowding and Galactic absorption are severe. Observations at longer wavelengths are needed to discover new background galaxies and complete the census in the ZOA.
    } 
   {The goal of this work is to identify galaxies behind the Milky Way Bulge using near-infrared (NIR) data from the VISTA Variables in Vía Láctea (VVV) survey.}
   {To this end we made use of different VISTA Science Archive (VSA) tools in order to extract relevant information from more than 32 billion catalogued sources in the VVV Bulge region. We find that initial  photometric restriction on sources from VSA \texttt{vvvSource} table combined with restrictions on star--galaxy separation parameters obtained from Source Extractor is a successful strategy to achieve acceptable levels of contamination (60\%) and a high completeness (75\%) in the construction of a galaxy target sample. To decontaminate the initial target sample from Galactic sources our methodology also incorporates the  visual inspection of false colour RGB images, a crucial quality control  which was carried out following a specifically defined procedure.}
   { Under this methodology we find 14480 galaxy candidates in the VVV Bulge region making this sample the largest catalogue to date in the ZOA.  Moreover these new sources provide a fresh picture of the Universe hidden behind the curtain of stars, dust and gas in the unexplored Milky Way Bulge region.}
   {The results from this work further demonstrate the potential of the VVV/VVVX survey to find and study a large number of galaxies and extragalactic structures obscured by the Milky Way, enlightening our knowledge of the Universe in this challenging and impressive region of the sky.}


   \maketitle
%

\section{Introduction}
\label{intro}

In the so--called Zone of Avoidance (ZOA) the extragalactic source detections beyond the Milky Way are hampered by Galactic absorption and stellar crowding. The dust and stars in our Galaxy prevent optical observations generating, as a consequence, an incomplete picture of the Universe behind this area of the sky. 

This  lack of information of the existing galaxies in the ZOA was firstly defined by \cite{Shapley61}. Since the ZOA covers about 25\% of the sky distribution of optically visible galaxies \citep{kra00}, several efforts have been developed to obtain larger information in different wavelengths.
In the optical wavelengths, the catalogues of \cite{kra00} and \cite{wou04} have allowed the detection of new galaxies at low Galactic latitude, although the gathering of information is restricted by Galactic dust and stars. On the other hand, near-infrared (NIR), X-ray and \ion{H}{I} radio surveys \citep{Roman1998, ebe02, Vauglin2002, kor04, pat05, skr06, Huchra2012} have detected galaxies and galaxy clusters at low Galactic latitude.
Also, \cite{Jarrett2000} have identified and extracted  extended sources from the Two Micron All-Sky Survey (2MASS) catalogue and \cite{mac19} presented redshifts for 1041 2MASS Redshift Survey galaxies, that previously lacked this information, mostly located within the Zone of Avoidance.
Furthermore, using the \ion{H}{I}  Parkes All-Sky Survey, \cite{sta16} have observed 883 galaxies, allowing to delineate the existence of possible large scale structure in the Great Attractor region, at low Galactic latitude. In addition, \cite{schr19} published a catalogue with 170 galaxies, from the blind \ion{H}{I} survey with the Effelsberg 100m radio telescope, located in the northern region of the ZOA. These results involve a great advance although a homogeneous coverage of the region has not yet developed, still leaving a large unexplored area. 

The near infrared public survey VISTA Variables in V\'ia L\'actea \citep[VVV,][]{min10,sai12} has proven that although the main scientific goals of VVV are related to stellar sources \citep{min11,bea13,iva14}, its exquisite depth (about 3 magnitudes deeper than 2MASS) and angular resolution (0.339 arcsec/px) make it an excellent tool to find and study extragalactic objects in the ZOA. The VVV photometry is divided into two different regions: the Disk and the Bulge. 
Within the VVV Disk, several works have been done using VVV data to identify galaxy candidates. In a pionering work, \cite{amo12} identified 204 new galaxy candidates from the VVV photometry of a 1.636 square degree region near the Galactic plane, increasing by more than an order of magnitude the surface density of known galaxies behind the Milky Way. 
Further, \cite{bar18} found 530 new galaxy candidates in two tiles in the region of the Galactic disk using a combination of SExtractor and PSFEx techniques  to detect and characterise these candidates. Later, \cite{bar19} confirmed the existence of the first galaxy cluster, discovered by the VVV survey beyond the galactic disk, by using spectroscopic data from Flamingos-2 at Gemini South Observatory. Moreover, \cite{bar21} studied the entire VVV Disk region discovering more than 5000 visually confirmed galaxies, of which only 45 were previously known, generating a VVV NIR galaxy catalogue of this region.

In the VVV Bulge, \cite{col14} found the VVV NIR galaxy counterparts of a new cluster of galaxies at redshift $z = 0.13$ observed in X-ray by SUZAKU \citep{mor13} detecting 15 new candidate galaxy members, within the central region of the cluster up to 350 kpc from the X-ray peak emission.
Further, \cite{gal21} (hereafter G21) found an unusual concentration of galaxies by exploring the $b204$ VVV Bulge tile, detecting 624 extended sources where 607 correspond to new galaxy candidates that have been catalogued for the first time. A significant overdensity of galaxies was detected in this region and the existence of a new galaxy cluster was confirmed by \cite{gal23} analysing photometric properties and estimating spectroscopic and photometric redshifts of the galaxy member candidates. Also, using Bulge VVVX data \citep{Minniti2016} \cite{gal22} present an IR view of Ophiuchus, the second brightest galaxy cluster in the X-ray sky,  finding seven times more cluster galaxy candidates than the number of reported Ophiuchus galaxies in previous works. 

All these works demonstrate the potential of the VVV survey in the detection and analysis of extragalactic sources in specific regions of the Galactic Bulge. Therefore it is crucial to define selection criteria suitable to identify extragalactic objects in the entire VVV Bulge region. 
Nevertheless, the identification of extragalactic sources in this area of the sky is really complicated, not only because of the high levels of stellar crowding and extinction but also given the huge amount of data compiled by the VVV survey. Therefore, our goal of identify galaxies behind the VVV Bulge is a data mining project that should extract, process and analyse relevant information from more than 590 million catalogued sources. In this work we accepted the challenge and present a procedure to select galaxy candidates hidden behind the curtain of stars, dust and gas in the unexplored VVV Bulge region.

This paper is structured as follows:  In section \ref{data}  we describe the main properties of the VVV data used in this work. In section \ref{targets_selection} we present a strategy to select extended objects from VVV data and we present the galaxy target sample in section \ref{targetsSE}. Our target selection criteria is tested in section \ref{testing} where we use public galaxy catalogues.  In section \ref{catalogue} we describe the construction of a galaxy catalogue in the  VVV Bulge sky region and finally in section \ref{conc} we summarise our main results and conclusions. 
The adopted cosmology throughout this paper is $\Omega = 0.3$, $\Omega_{\Lambda} = 0.7$, and $H_0 = 100 \rm kms^{-1} \rm Mpc$.

\section{VVV data}
\label{data}
The VVV survey is a NIR ESO (European Southern Observatory) public survey set up to map the Milky Way bulge and disk. The survey uses the 4.1-m ESO Visual and Infrared Survey Telescope for Astronomy \citep[VISTA,][]{Emerson2004,Emerson2006,Emerson2010} located at ESO’s Cerro Paranal Observatory in Chile. 
Observations were carried out with the VISTA IR CAMera (VIRCAM), which is composed of 16 Raytheon VIRGO 2048x2048 HgCdTe on CdZnTe substrate science detectors, with a mean pixel scale of 0.339 arcsec px$^{-1}$ and a field of view per exposure of 0.59 square degrees. The detectors are arranged in a grid with spaces in between detectors of 90$\%$ of the detector width in the x-direction and 42.5$\%$ in the y-direction. Six exposures are required to make a contiguous area of 1.636 square degrees. The individual exposure is known as a `pawprint' and the final area as a `tile' (1.48 by 1.18 degrees in size). 

The Bulge of the Milky Way, located at nearly 8 kpc from the Sun, is a challenging region where the optical light from millions of stars at low latitudes in our Galaxy is hidden by dust and gas. The VVV survey has mapped the Bulge area between $-10\deg<l<+10.4\deg$ and $-10.3\deg <b<+5.1\deg$ by 196 contiguous tile images covering about 320 square degrees using Z, Y, J, H and Ks near-IR filters. The observing strategy involves adding some overlap between tiles for a smooth match, therefore there are overlaps between adjacent tiles and the complete survey area is covered by at least four exposures in each filter. These observations  overcome the heavy extinction of the Galactic Plane reaching  limiting aperture magnitude up to Ks$\sim$18 in clean fields and Ks$\sim$16 in the inner bulge \citep{min10,sai12}.

The data reduction consists of two main steps, first the basic data reduction was carried out by VISTA Data Flow System \citep[VDFS,][]{Emerson2004,Irwin2004,Hambly2004} at the Cambridge Astronomy Survey Unit (CASU\footnote{http://casu.ast.cam.ac.uk/surveys-projects/vista}). Then a second order data processing involves the production of survey products as calibrated data and catalogues and  was done by the Wide Field Astronomy Unit’s (WFAU) VISTA Science Archive (VSA\footnote{http://horus.roe.ac.uk/vsa/index.html}) in Edinburgh.  

 The VSA is therefore a virtual observatory that holds the image and catalogue data products generated by the six VISTA Public Surveys, including VVV. It is a relational database that stores information of catalogued sources available through Structured Query Language (SQL) queries. Data are presented in tables which are linked via reference ID numbers. These tables contain catalogues with main information of astronomical objects and also the metadata and calibration image information. Also, VISTA images are held in the archive as multi-extension FITS files and different tools at VSA can be used to  extract cut-out images around a given position. 
 
 In the present work we made use of different VSA tools in order to extract relevant information of particular sources in the VVV Bulge region. To this end we have considered the latest VSA release of the VVV survey, VVV-DR5.

\section{Target selection strategy}
\label{targets_selection}
 As principal catalogue we use VSA \texttt{vvvSource} table that contains merged records from the deepest observations in each detection for a given object in the five ZYJHKs VVV passbands. In VSA the source tables combines single-passband  detections into a merged multicolour record, propagating only the most-useful subset of photometric, astrometric and morphological attributes along with associated errors. Also, the source table presents a normally distributed merged classification statistic that describes the probability of an object to be point-like compared to an empirical idealised model based representing the PSF for the frame. A value of 0.0 corresponds to point-like sources, increasingly negative values indicate sharper images (e.g. noise-like) and increasingly positive values indicate extended sources as resolved galaxies \citep[for a detailed description see][]{Hambly2008}.

 \begin{figure}
  \centering
   \sidecaption
   \includegraphics[width=0.45\textwidth]{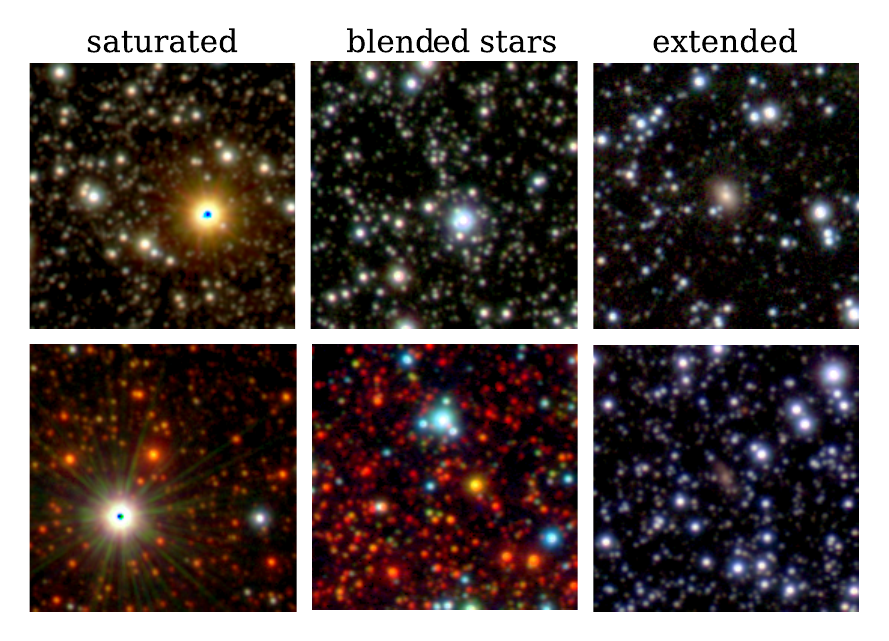}
      \caption{Examples of the different classes of objects identified in the VVV Bulge region, saturated sources (left), blended stars (middle) and extended objects (right). The RGB images are a combination of KsJZ filters. Each image has a dimension of 1 arcmin x 1 arcmin.
            }
         \label{types_examples}
  \end{figure}

Even if the source table comprises only the most useful information of a detected source, the \texttt{vvvSource} table contains more than 300 entries for each of the  $\sim$ 3.7 billion catalogued sources (10<Ks<17). From these, 1.2 billion objects have a higher probability of being galaxies according to the source--type classifications, i.e. \textit{mergedClass=1}. This number is huge compared to the expected number of galaxies to be found in the VVV Bulge calculated in G21 using  mock catalogues, where we study VVV Bulge b204 tile, finding that for a  mean background region the integrated number of galaxies per 1 deg$^2$ is about 70 galaxies with Ks$<$17 up to z$\sim$0.15. If we consider the entire VVV Bulge region (320 deg$^2$) the expected number of galaxies based on these results is 22400. It is worth to notice that the absorption and stellar density in b204 tile is low compared to another bulge regions and therefore the number of galaxies may vary according to the tile position in this complex area of the sky.

Therefore, it is crucial to define suitable constraints in the main photometric parameters of the catalogued objects in the \texttt{vvvSource} table, in order to debug this sample from non--extragalactic sources.

In order to perform a suitable selection of targets with a higher probability of being galaxies, we use SQL queries on \texttt{vvvSource} table taking into account that galaxies present particular photometric properties, such as red colours, that can be used to differentiate extended galaxy--like objects from stellar sources and also use specific star--galaxy separation parameters that help in the identification of galaxy--type objects.

In order to select galaxy candidates we consider the merged class statistic limiting our study to galaxy--type classification \textit{mergedClass=1} and rejecting all other source-types (e.g.  noise, star, probable stars and probable galaxies).
To select photometric restrictions we use the non-aperture corrected magnitudes within 2.0 arcsec diameter (APERMAG3) and consider the \cite{Chen2013} maps  in the \citet{nis09} system (stored in \texttt{vvvBulge3DExtinctVals} and \texttt{FilterExtinctionCoefficients} EXTINCT tables) to obtain extinction corrected magnitudes.

We require available photometry (not null values) in the JHKs bands and restrict our selection to sources with 10 mag$<$Ks$<$16.2 mag. The lower limit is chosen to avoid saturated stars and the upper limit is close to the limiting magnitude in the inner bulge \citep{sai12} and is also the reliable limit for visual inspection defined in G21. Different works on VVV data show that there are no galaxies with (J-Ks) colours lower than $\sim$0.5 mag \citep[e.g.][]{amo12,col14,bar18,gal21}. Therefore as first restriction we  consider J-Ks$>$0.5 mag, H-Ks$>$0 mag, J-H$>$0 mag colour constrains in the selection of probable galaxy--type sources.

Under these photometric restrictions we obtain an initial sample of  about 7.5 million sources in the whole VVV Bulge region. To study the nature of the selected sources we build a random sample comprising 2000 objects randomly selected from the initial sample of 7.5 million sources.
With this sample we  perform a detailed study aiming to depurate the original sample from galactic sources while retaining most of the galaxy-type objects.

\subsection{Different source classes in the VVV Bulge}
\label{sources-types}

Attempting to identify galaxies, we consider several restrictions on VVV catalogued sources. Nevertheless it was initially expected a high level of contamination in our sample given the complexity of the galactic Bulge. 
The visual inspection is then an unavoidable step in the identification of galaxies in VVV images \citep{amo12,bar18,gal21}. 
We test different filter combinations in order to highlight galaxy--like colours compared to stellar galactic sources and find that building RGB images from KsJZ filters is the most convenient combination. This is because the blue light of galaxies behind the galactic bulge is absorbed by the Milky Way dust, making galaxies appear as red objects compared to the bluer stellar sources.

In orden to study the nature of the selected sources we perform a visual inspection of RGB images of the objects in the random sample and identify 3 main different types of sources:  blended stars, saturated sources and extended objects. In Fig. \ref{types_examples} we show examples of these three different classifications.

\begin{figure}
   \sidecaption
   \centering
   \includegraphics[width=0.45\textwidth]{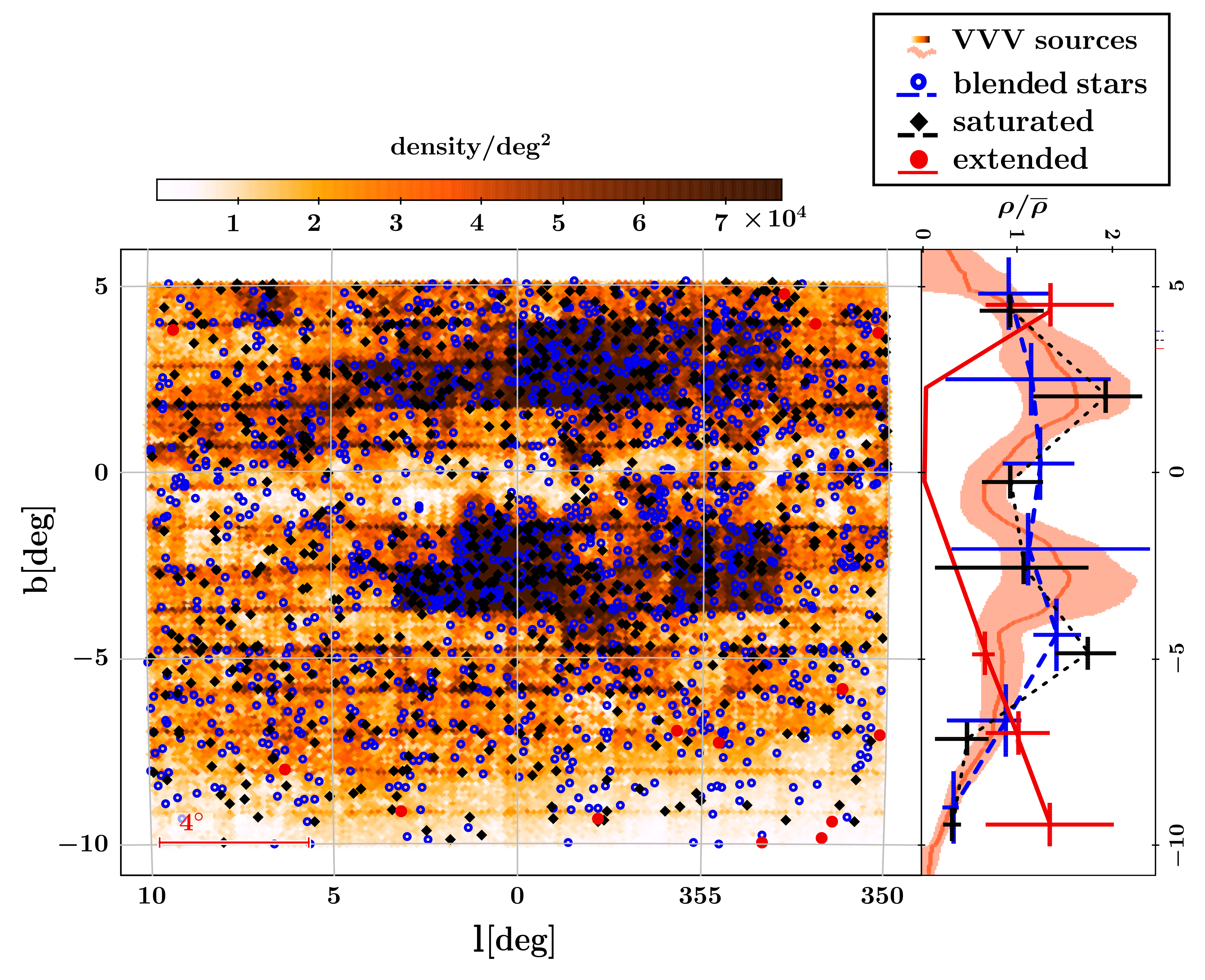}
      \caption{ Sky distribution of the 7.5 million  sources selected from VSA VVV-DR5 catalogue considering \textit{mergedClass=1}, 10 mag$<$Ks$<$16.2 mag, J-Ks$>$0.5 mag, H-Ks$>$0 mag, J-H$>$0 mag). We pixelate the VVV Bulge region in pixels of 7 arcmin x 7 arcmin, colour codded according to the density per square degree. We also show the distribution of extended objects, saturated sources and blended stars in a sample of 2000 random sources. The right panel shows the overdensity as a function of galactic latitude for the entire sample and also for the different classes of sources in the random sample. Error bars correspond to 1$\sigma$ values.
              }
         \label{skyplot_2000random}
  \end{figure}

As first result we found that the 99.3\% of the sources are non-extended with 775 saturated sources and 1211 blended stars (38.75\% and 60.55\%, respectively). We found only 14 (0.07\%) sources with the extended category, therefore the random sample is highly dominated by sources associated with our Galaxy  
In Fig. \ref{skyplot_2000random} we show the sky distribution of objects in the random sample over-imposed on the sky density of the initial sample of 7.5 million sources. We calculate the overdensity as a function of galactic latitude and find that the initial sample is severely depeated at low galactic latitudes ($|b|<2$). Saturated sources and blended stars also show this trend. 
There is a lack of extended objects at these low  galactic latitudes, so it is expected that 99.9\% of sources at low latitudes are associated with objects in our Galaxy. These results are not surprising given the high absorption and stellar density of the inner bulge region.

\subsection{Photometry requirements}
\label{mag_restrictions}
\begin{figure}
   \sidecaption
   \centering
   \includegraphics[width=0.45\textwidth]{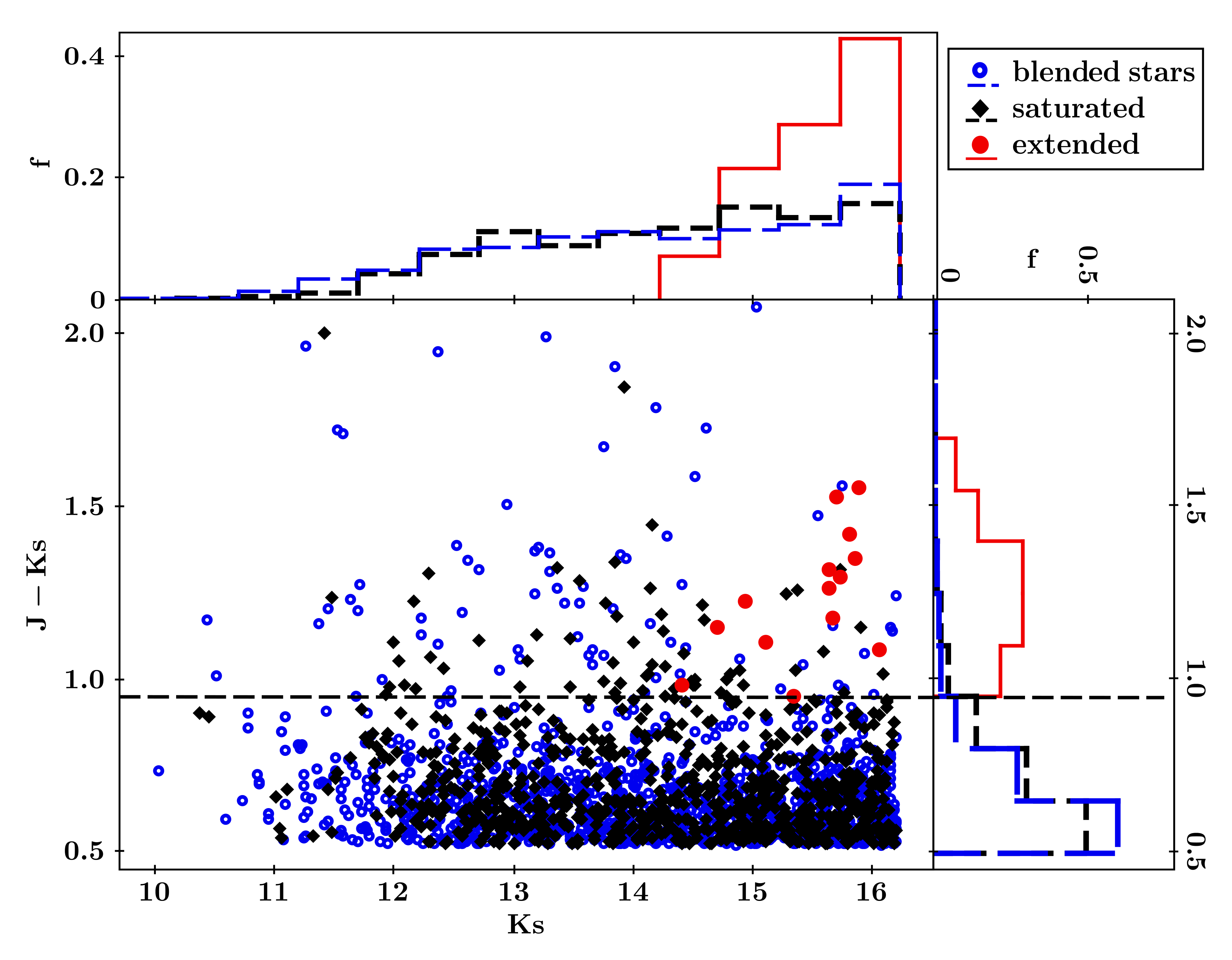}
      \caption{Colour--magnitude diagrams (J-Ks) versus Ks for sources in the random sample considering extinction corrected aperture magnitudes. We present saturated sources as black diamonds, blended stars as blue open circles and extended objects as red dots. Long-dashed lines correspond to the (J-Ks)=0.95 mag colour cut adopted in this work. In the top and right panels we plot the normalised distributions of colour and magnitude, respectively (short dashed line for saturated sources, long dashed line for blended stars and red solid line for extended objects).
              }
         \label{CM_randomi}
  \end{figure}

In this section we study different photometric properties of objects in the random sample as colours and magnitudes, considering the different sources types defined previously. 
We plot in Fig. \ref{CM_randomi} (J-Ks) versus Ks colour magnitude diagram. From this plot we can clearly appreciate that extended objects are redder than (J-Ks)=0.95 mag. Moreover only 132 (6.6\%) of the non-extended sources (71 blended stars and 61 saturated sources) present (J-Ks) colours as red as extended galaxy candidates. Most of these sources are associated with red stars at low galactic latitudes ($|b|<2$). Therefore we adopt (J-Ks)>0.95 mag as a suitable colour limit to further decontaminate  our sample from stars. From Fig. \ref{CM_randomi} we also see that galaxies are fainter than Ks=14 mag nevertheless different authors report galaxies with Ks magnitude as bright Ks=10 mag within the VVV survey \citep[see G21 and][]{bar18}. Giving the high level of decontamination archived with the (J-Ks) colour cut, we decide to explore the entire 10 mag $<$Ks$<$16.2 mag range.

We also find that 12 of the 14 (86\%) of the detected extended sources in the random sample have available photometry in the five ZYHJKs bands. 
On the other hand,  only 56 (29 saturated sources and 27 blended stars) of the 132 (42\%)  non--extended sources with (J-Ks)>0.95 mag have available photometry in all VVV bands. Therefore, we conclude that by requiring ZYHJKs with available photometric measurements we will be able to discard most galactic sources at low latitude, while retaining about 90\% of extended objects in the VVV Bulge region. 

\subsection{Obtaining suitable photometry of extended sources}
\label{SEcuts}
The catalogues obtained from VSA have only aperture magnitudes, but for galaxies it is crucial to also estimate total magnitudes. Therefore in order to obtain suitable photometry of  extended sources we run SExtractor \citep[SE, ][]{ber96} to create new catalogues from VVV images. 
The SE software was developed to detect, deblend, measure and classify sources from astronomical images. As crowded fields are very common in the VVV sky area, the deblending of SE is very useful to analyse overlapping objects.

We use SE with parameters setting according to G21. Briefly SE recognised and separated from the background noise, all the objects with a threshold more than twice the medium brightness of the sky and spanned over at least ten connected pixels on the VVV images \citep{col14,bar18,gal21}. A Gaussian filter, with 3 x 3 pixels with a convolution mask was applied on the images in order to efficiently separate low-surface-brightness objects from spurious detections. For the deblending process we consider the default values that allow overlapped stars to be efficiently separated from galaxies in this tile. We use zero point magnitudes to calibrate the magnitudes estimated by SE.
The output parameters selected from SExtractor catalogues were equatorial coordinates, Ks aperture magnitude calculated in three-pixel-radius circular aperture, Ks total magnitude  MAG$\_$AUTO, which is based on Kron's algorithm \citep{kro80},  CLASS\_STAR (CS) and half$-$light radius ($\rm R_{1/2}$). These last two parameters are used for star-galaxy separation, since the stars lie in a sequence near CS = 1 while galaxies are near CS = 0 and the two sequences merge for faint sources. The half-light radius parameter, $\rm R_{1/2}$ measures the radius that encloses 50\% of the object total flux. For objects larger than the seeing, $\rm R_{1/2}$ is independent of magnitude and is larger for galaxies than for stars.

As described previously, the restrictions on the photometry are not enough to properly select galaxies from the \texttt{vvvSource} table because there is a large fraction of targets associated with stellar sources in the Galaxy. In this sense we expect that  SE do the job and help us to lower the contamination in our sample, while keeping a high completeness level. To this end we use the \texttt{MultiGetImage} tool from VSA  to download 1 arcmin x 1 arcmin images in the Ks band countered in the random sources. 

We run SE on these images finding that 1027 sources were successfully extracted, (11 extended, 238 saturated and 778 blended stars) from the original list of 14 extended objects, 775 saturated sources and 1211 blended stars. Therefore SE  identifies about 80\% of extended objects, discards a large fraction (70\%) of saturated sources but removes only 36\% of blended stars. 

\begin{figure}
   \sidecaption
   \includegraphics[width=0.49\textwidth]{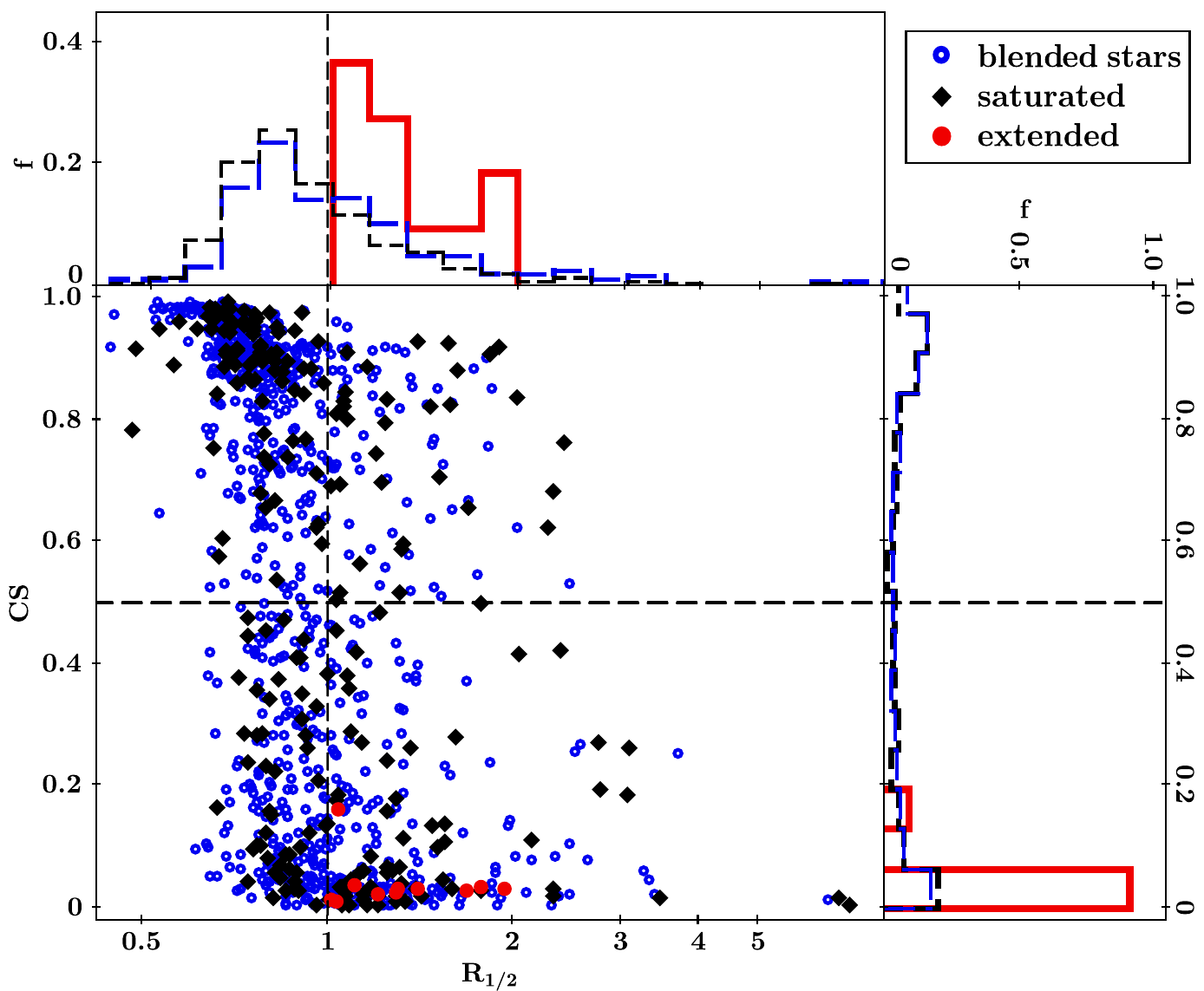}
      \caption{Distribution of $\rm R_{1/2}$ (in arcsec) versus CS for sources in the random sample. We present saturated sources as black diamonds, blended stars as blue open circles and extended objects as red dots. Long-dashed lines correspond to the  CS and $\rm R_{1/2}$ cuts adopted in this work. In the top and right panels we plot the normalised distributions of $\rm R_{1/2}$ and CS, respectively (short dashed line for saturated sources, long dashed line for blended stars and red solid line for extended objects).
              }         
         \label{CSR_randomi}
  \end{figure}

To decontaminate from stellar sources we study the star-galaxy separation parameters CS and $\rm R_{1/2}$. In Fig. \ref{CSR_randomi} we present a dispersion plot of these parameters and also their individual distribution. From this figure it can be appreciated that most extended sources present a low CS parameter value, with all the SE detected extended objects in the random sample having CS$<$0.2. In the case of both, saturated sources an blended stars, the CS distribution is bimodal with about half of the sources presenting CS$<$0.5. For the $\rm R_{1/2}$ parameter there is a clear difference between the distribution of extended objects compared to saturated sources and blended stars. For instance, if we consider $\rm R_{1/2}\sim 1$ arcsec we find that about the 75\% of stellar sources associated with our Galaxy have values lower than this limit, while all the extended objects present  $\rm R_{1/2}>$1 arcsec.

Based on these results we consider CS<0.5 and $\rm R_{1/2}>$1 arcsec restrictions as suitable cuts in the star--galaxy separation parameters. From the original random sample we find that 11 extended objects, 67 saturated sources and 164 blended stars fulfil these restrictions. Therefore, considering only restrictions on the SE star--galaxy separation parameters, the contamination associated with Galactic sources is lowered to a $\sim$20\% level while completeness is preserved at levels closer to 80\%.

\section{A galaxy target sample in the VVV Bulge }
\label{targetsSE}

\begin{table*}[htb]
\centering
\caption{Numbers of extended objects, saturated sources and blended stars in the random sample (\textit{mergedClass=1}, 10 mag$<$Ks$<$16.2 mag, J-Ks$>$0.5 mag, H-Ks$>$0 mag, J-H$>$0 mag), considering different restrictions in VVV photometry and SE star--galaxy separation parameters.}
\begin{tabular}{|l c c c c| }
\hline
random & Extended objects  & Saturated sources & blended stars & total    \\
\hline
\hline
All             & 14 & 775  & 1211 & 2000  \\
(J-Ks)>0.95 mag    & 14 & 61  & 71 &  146  \\
(J-Ks)>0.95 mag and APERMAG3>0 mag & 12 & 29  & 27 &  68 \\
CS<0.5 and $\rm R_{1/2}>$1 arcsec & 11  & 67 & 164 & 242  \\
(J-Ks)>0.95 mag, APERMAG3>0 mag, CS<0.5 and $\rm R_{1/2}>$1 arcsec   & 11  & 5 & 14 & 30 \\
\hline  
\end{tabular}
\label{tt}
\end{table*}

\begin{figure*}[]
    \centering
       \includegraphics[width=0.9\textwidth]{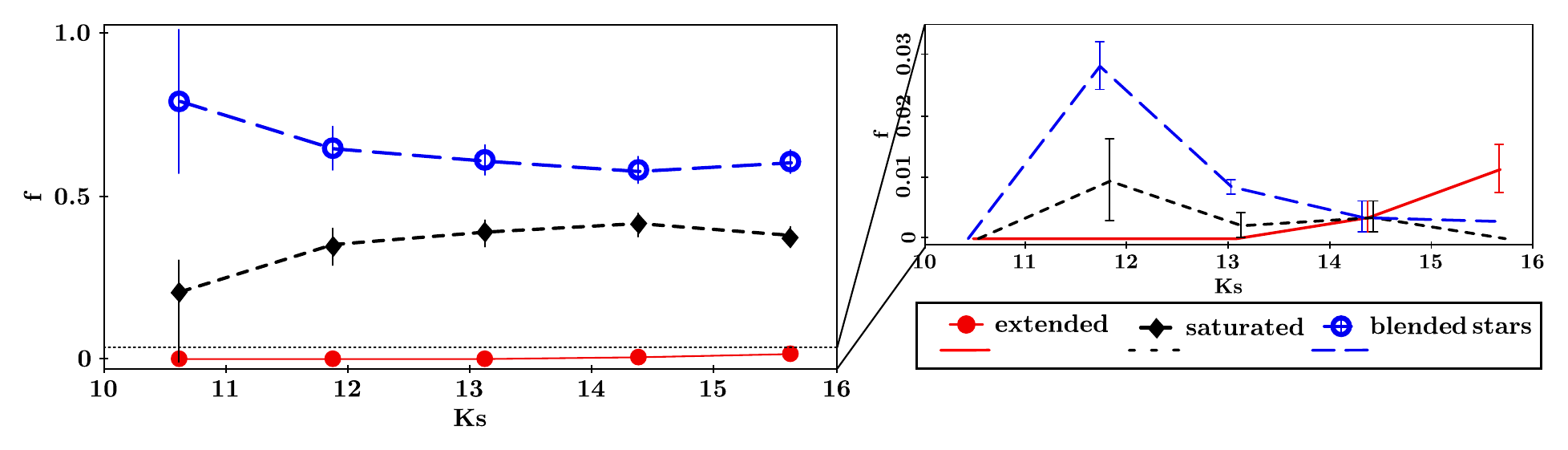}

    \caption{Fraction of extended objects (red solid line), saturated sources (black short--dashed line) and blended stars (blue long--dashed line) in the random sample as a function of Ks extinction corrected aperture magnitude. In the left panel we plot the original fractions and in the right panel we show the fractions after considering the final restrictions. Error bars represent the standard error of the fractions.}
    \label{FclassKs}
\end{figure*}

Based in the results of the previous sections, we assume that initial  photometric restrictions on sources from VSA vvvSource table, combined with restrictions on star--galaxy separation parameters obtained from SE, seem to be a successful strategy to achieve acceptable levels of contamination and a high completeness  in the construction of a sample suitable for the identification of galaxy candidates in the VVV Bulge.

Therefore we define the following final restrictions:

\begin{itemize}
    \item \textit{mergedClass}=1
    \item ZYJHKs APERMAG3 magnitudes with not null values
    \item Ks APERMAG3 extinction corrected aperture magnitude in the range  10 mag$<$Ks$<$16.2 mag
    \item J-Ks$>$0.95 mag, H-Ks$>$0 mag, J-H$>$0 mag considering  extinction corrected aperture magnitudes
    \item SE star--galaxy separation parameters obtained from Ks images fulfilling CS<0.5 and $\rm R_{1/2}>$1 arcsec.
\end{itemize}

We assume these restrictions in our intent to minimise contamination in our sample, while retaining most of the possible galaxy candidates. From the 2000 sources in the random sample, there are only 20 that pass these cuts, 11 extended objects, 5 saturated sources and 14 blended stars. Therefore we expect that one third of the objects under study are galaxy candidates. Table \ref{tt} summarises the number of extended objects, saturated sources and blended stars for each tested restriction. From these numbers it can be appreciated that combining photometric and SE restrictions we lower the contamination from 99\% to 60\% meanwhile we retain about 80\% of the extended sources.

In Fig. \ref{FclassKs} we show  the fractions of extended objects, saturated sources and blended stars in the random sample as a function of Ks extinction corrected aperture magnitude. We study the trend before and after considering our final restrictions. 
An increment of extended objects can be seen throughout fainter magnitudes with galaxy--like sources having Ks$<14$ mag. Also, the fraction of blended stars in the sample drops for fainter magnitudes, while the fraction of saturated sources slightly increases for sources fainter than Ks=13 mag.  After the restrictions in the photometry and in the star--galaxy separation parameters (obtained from SE), the fraction of galactic sources significantly drops and is important only at intermediate magnitudes. In the case of the extended sources the trend remains unaltered after the photometric plus SE restrictions.

From the photometric restrictions we get a sample of 191879 VSA sources, and after running SE we retain 38718 targetsSE, to be analysed in the search of galaxy candidates behind the VVV galactic bulge region. Note that by running SE on Ks images we  expect to avoid the visual inspection of about the 80\% of the sources selected only with the photometric restrictions, which significantly helps in this very time consuming but crucial step of our methodology.

\section{Testing target selection}
\label{testing}
In the previous section we define selection criteria for extended objects in the VVV Bulge region, aiming to decontaminate the catalogue of VVV sources from saturated/blended Galactic stars.  
The aim of this Section is to test our target selection criteria in the identification of already catalogued galaxies in the VVV Bulge region. To this end we will use the 2MASS Extended Source Catalogue \citep{Jarrett2000} as well as the catalogue of galaxies candidates presented in G21, to test the completeness of these samples under our galaxy selection criteria.

\subsection{Comparison with  2MASS Extended Source Catalogue}
\label{c2m}
\begin{figure}
   \sidecaption
   \includegraphics[width=0.45\textwidth]{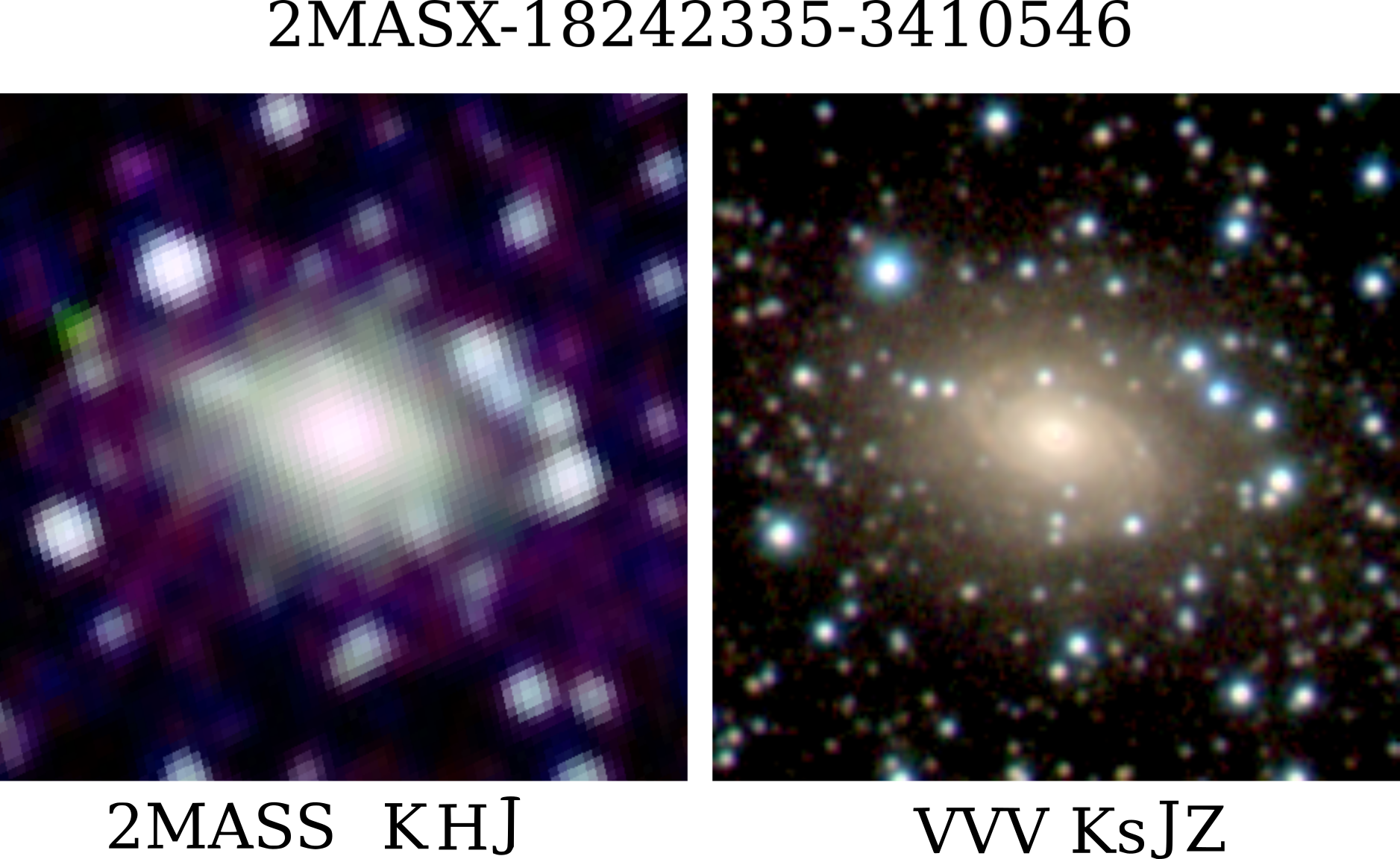}
      \caption{ RGB images, KHJ 2MASS (left) and KsJZ (right), of 2MASX-18242335-3410546 galaxy. The images size is 1 arcmin x 1 arcmin.  
              }
         \label{2mass_VVV_comparison}
  \end{figure}
  
The largest catalogue of extended sources in the VVV Bulge region, to date, is the 2MASS Extended Source Catalogue \citep[2MXSC][]{Jarrett2000}. This is an all--sky survey comprising more than 1.5 million objects (2MASX) but only 271 of these sources are in the VVV Bulge area considered in this work. 
In order to study in more detail these sources, we build 1 arcmin x 1 arcmin RGB images using VVV KsJZ bands which highlight the red colour of galaxies against the bluer foreground stars. This effect together with the fact that VVV is deeper than 2MASS and has a higher resolution, allows a detailed visual inspection of 2MASX objects leading to a suitable morphological classification as can be seen in Fig. \ref{2mass_VVV_comparison}, where the spiral nature of 2MASX-18242335-3410546  galaxy clearly emerges in the VVV false colour image. 

\begin{figure}
   \sidecaption
   \includegraphics[width=0.5\textwidth]{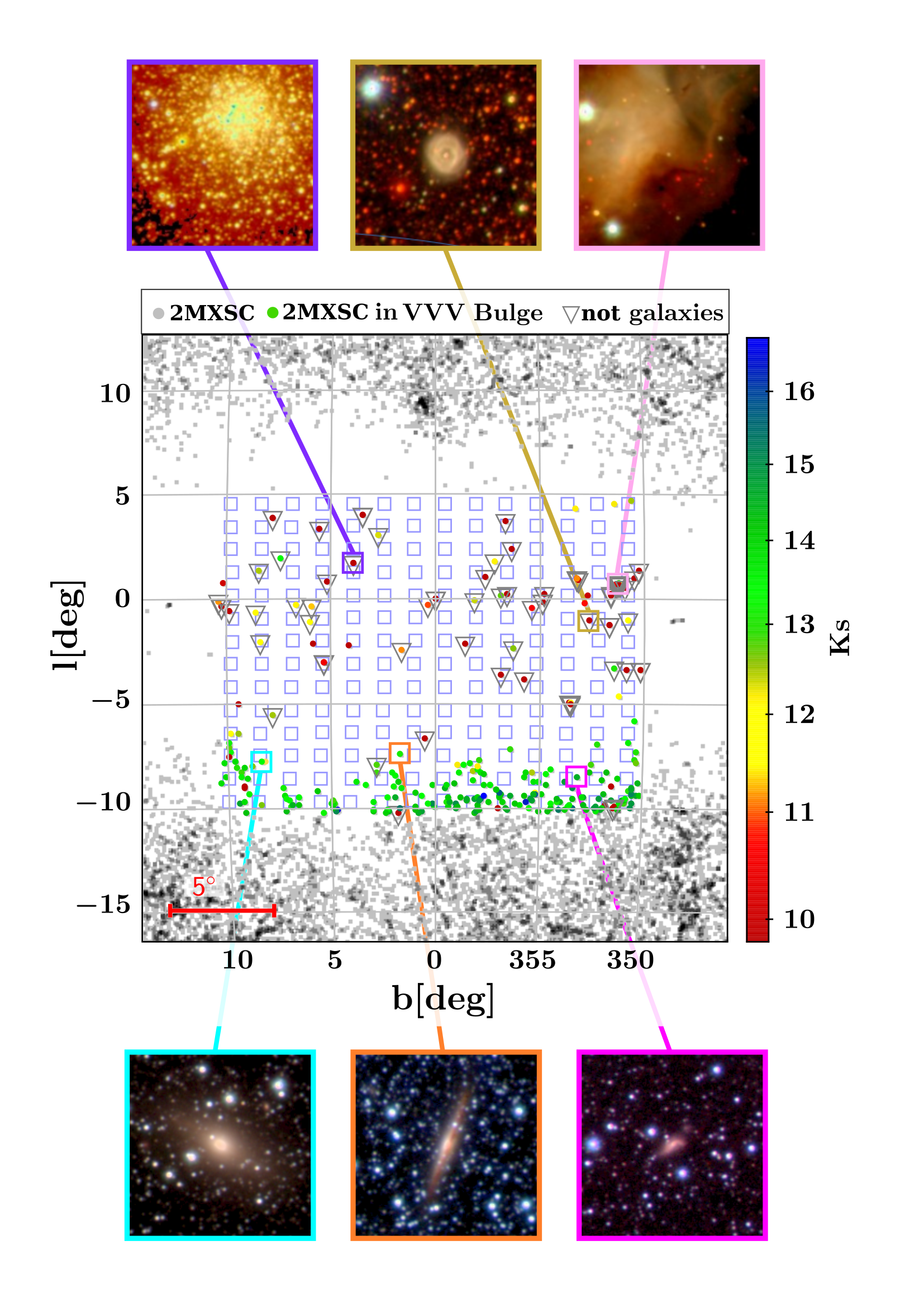}
      \caption{Sky distribution of objects in the VVV Bulge area. Cyan open squares represent the position of the VVV tiles. Grey dots are the 2MXSC sources and points represent 2MXSC in VVV Bulge (colour--codded according to Ks extinction corrected aperture magnitude). Open triangles represent extended objects that are not galaxies. In the top of the figure we show 3 RGB (KsJZ) images of 2MXSC objects associated with Galactic objects and in the bottom we show 3 examples of 2MXSC galaxies.
           }
         \label{examples2MXS}
  \end{figure}

We performed a careful visual inspection of the 271 2MXS in VVV Bulge finding that only 182 are galaxies, the remaining objects are associated with Galactic sources as star clusters and gaseous regions. In Fig. \ref{examples2MXS} we show the sky distribution of 2MASX sources in the VVV Bulge area and some examples of Galactic and extragalactic objects in this sample. The Galactic sources are the brightest in the sample (Ks$<$12 mag) and are located mainly at b$>$-5 deg. All galaxy--like objects present magnitudes fainter than Ks$\sim$ 12 mag and are located at latitudes lower than b=-7 deg.

\begin{figure}[]
    \centering
       \includegraphics[width=0.49\textwidth]{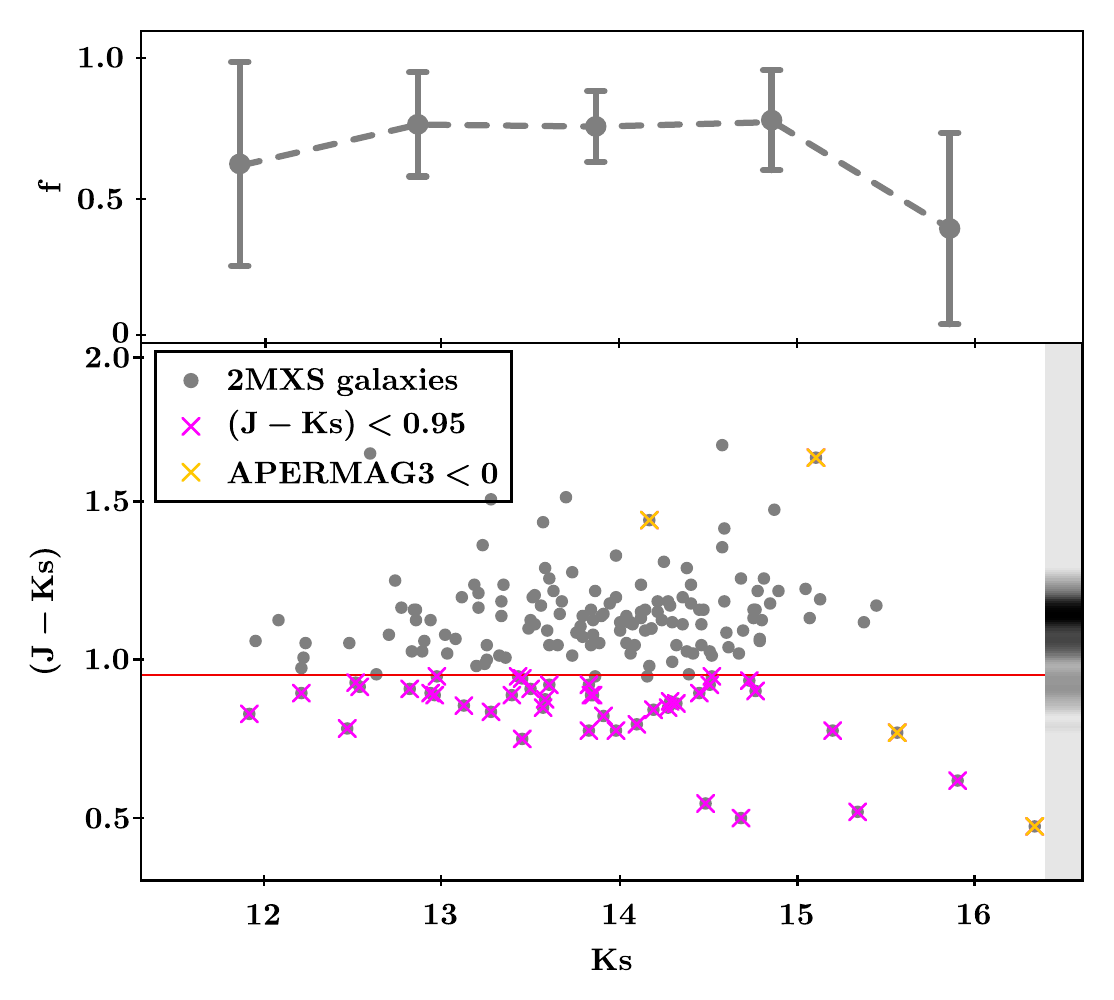}
    \caption{Colour--magnitude diagrams (J-Ks) vs Ks  of the 2MXSC galaxies (grey dots). We consider VVV extinction corrected aperture magnitudes.  Crosses show galaxies not detected under our methodology, in pink the sources lost by the (J-Ks) cut and in yellow two galaxies with unreliable measurements in Z and Y VVV bands. In the right region of this figure we plot a densogram based on the (J-Ks) colour distribution of the sources. In the upper panel we show the completeness based in 2MXS galaxies as a function of Ks magnitude. }
    \label{JKSKs2MXS}
\end{figure}

In our targetsSE list we find 136 out of the 182 2MXSC galaxies, therefore we are able to recover the 75\% of the catalogued 2MASS galaxies. From the 46 lost objects, 6 of them do not have magnitude measurements in all VVV bands and  40 present (J-Ks)$<$0.95 mag, our colour selection limit. In Figure \ref{JKSKs2MXS} we show (J-Ks) vs Ks colour--magnitude diagrams of the 2MXS galaxies under study. We consider extinction corrected aperture magnitudes. In this plot we show galaxies lost by the (J-Ks) cut and also 4 galaxies with unreliable measurements in Z and Y VVV bands. In the right region of this figure we plot a densogram based on the (J-Ks) colour distribution of the sources and as can be seen most 2MXSC galaxies are grouped around the  (J-Ks)=0.95 mag limit adopted in this work. In Fig. \ref{JKSKs2MXS} we also show the completeness based in 2MXSC galaxies as a function of Ks extinction corrected aperture magnitude. The completeness rate is almost constant, with differences only in the faintest bin where it drops from 75\% to 40\% .

\subsection{Comparison with G21 galaxy candidates}

VVV Bulge tile $b204$ has been analysed in G21 for the search of galaxy candidates. This tile is a region of $1.475 \times1.109$ deg$^2$ located at the edge of the VVV Bulge ($l=355.182^o$ and $b=-9.68974^o$). In G21 we present a catalogue of 624 galaxy candidates with total magnitudes in the range $10<Ks<17.5$ and only 18 of these objects have been previously catalogued.  

In this section we re--explore b204 tile under our selection criteria. From the original 624 catalogued sources in G21 there are 275 that are within the area covered by the extinction map used in this work  (no extinction values for b$<$-10.1 deg and b$>$5.1 deg) and have magnitudes 10 mag$<$Ks$<$16.2 mag, the range considered in G21 to select reliable galaxies (here after G21 galaxies). From these G21 galaxies, 212 are in our targetsSE list. We lose 18 sources with non available photometry in all VVV bands,  11 that do not fulfill the colour cut (J-Ks)$>$0.95 mag and 34 sources that do not pass SE selection given that in G21 we adopt a more flexible restriction $\rm R_{1/2}>$0.7 arcsec. Therefore we  recover the 77\% of G21 galaxies within the magnitude range under consideration.

\begin{figure}
   \sidecaption
   \includegraphics[width=0.49\textwidth]{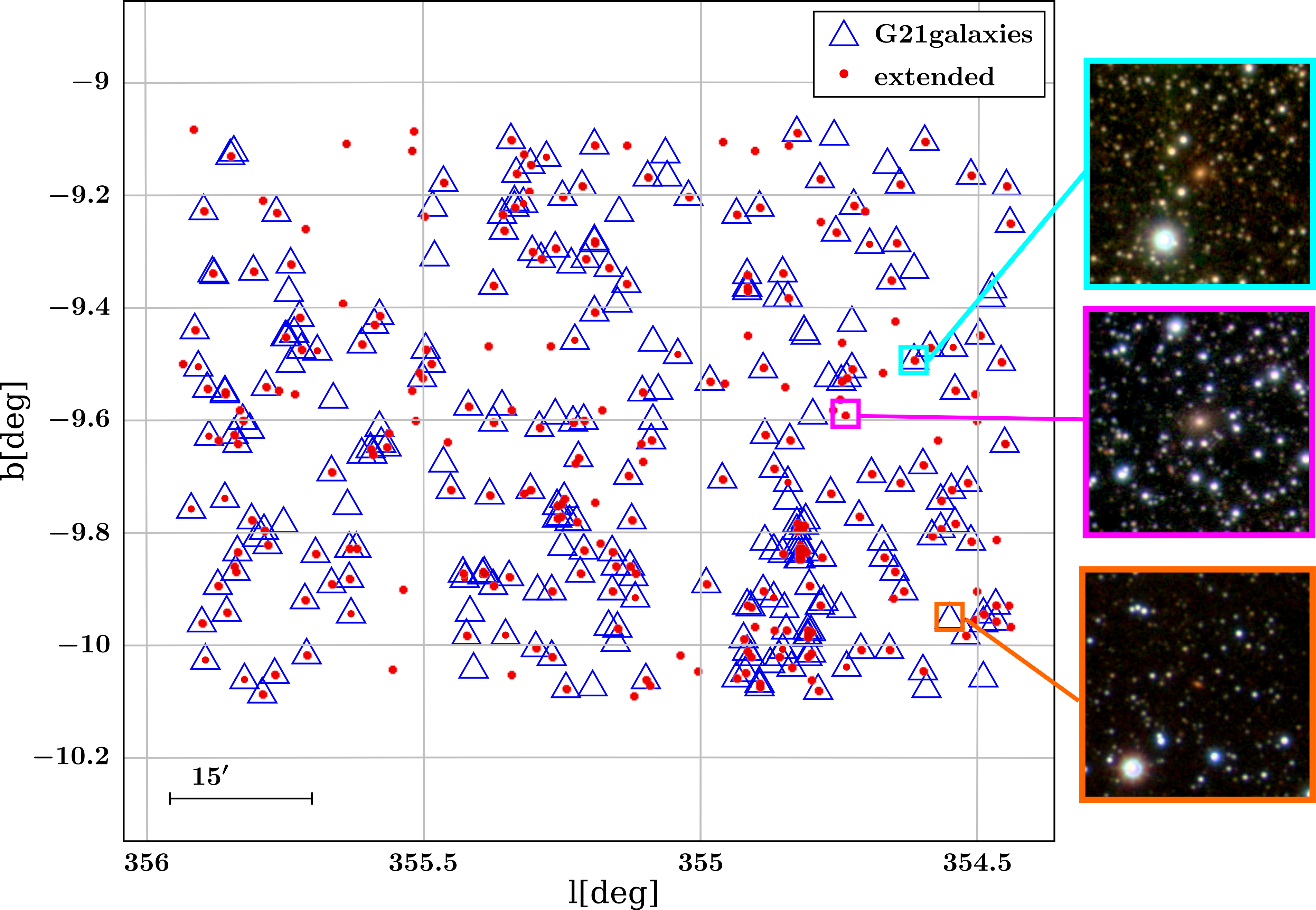}
      \caption{Sky distribution of G21 10 mag$<$Ks$<$16.2 mag galaxy candidates (blue triangles) and extended sources identified in this work within b204 tile (red dots). In the right region of the figure we show one example of G21 galaxy that match our extended source list (top), an extended source that has not been identified in G21 (middle) and a G21 galaxy that does not fulfill our extended selection criteria (bottom), respectively. The images are 1 arcmin x 1 arcmin.
              }         
         \label{skymapb204}
  \end{figure}

Moreover, under the methodology adopted in this work we select 88 new extended sources in b204 tile allowing the identification of a similar number of  10 mag$<$Ks$<$16.2 mag galaxy candidates compared to G21 methodology (300 this work 275 G21). In Fig. \ref{skymapb204} we plot the sky distribution of these sources.

 \begin{figure}
   \sidecaption
   \includegraphics[width=0.49\textwidth]{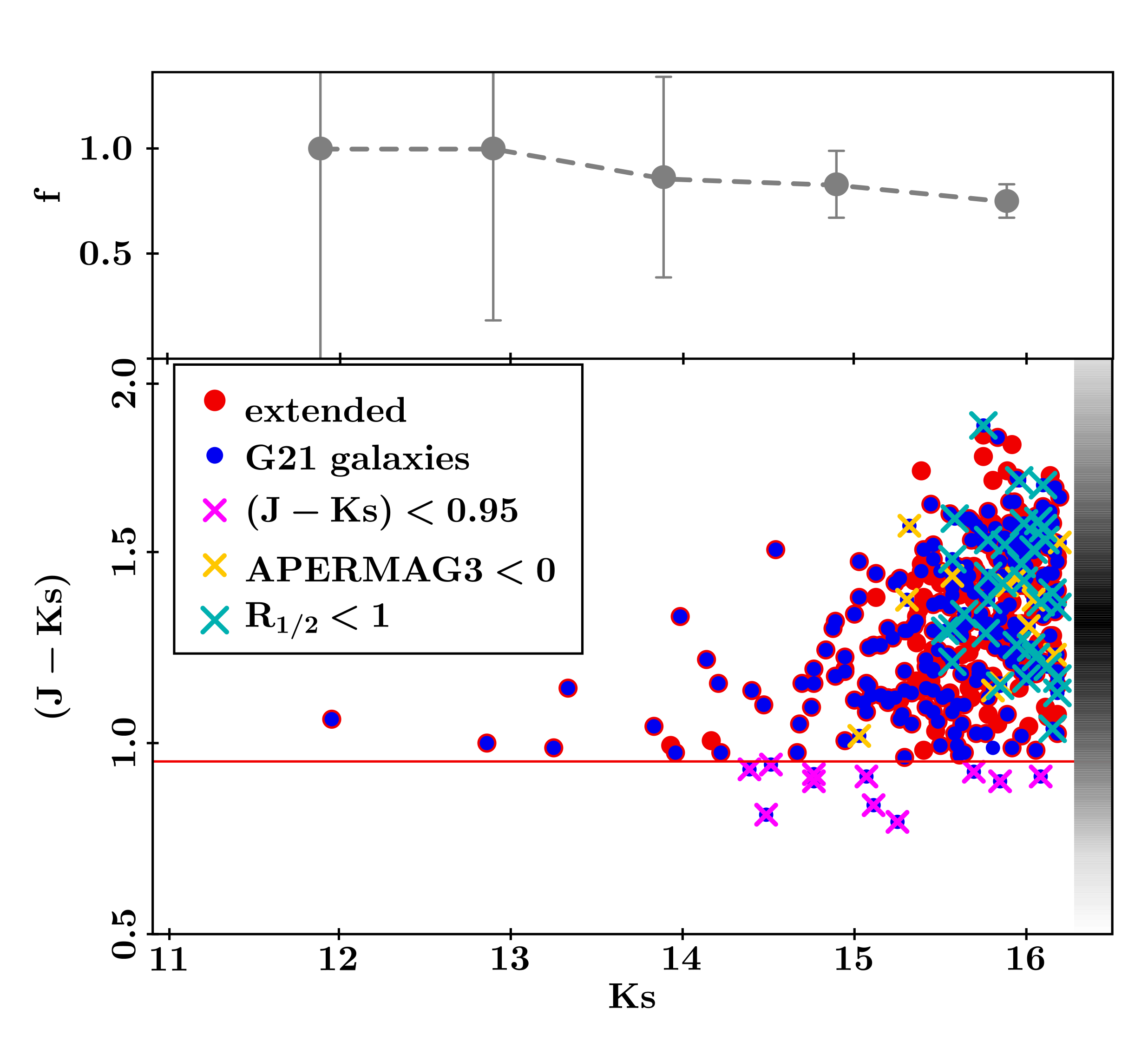}
      \caption{Colour--magnitude diagrams (J-Ks) vs Ks for the extended objects in $b204$ tile (red dots), G21 galaxies (blue dots). We consider VVV extinction corrected aperture magnitudes. Crosses show galaxies not detected under our methodology, in pink the sources lost by the (J-Ks) cut, cyan sources with $\rm R_{1/2}<$1 arcsec  and in yellow  galaxies with unreliable measurements in the Z and Y VVV bands. In the right region of this figure we plot a densogram based on the (J-Ks) colour distribution of the sources. In the upper panel we show the fraction of recovered G21 galaxies as a function of Ks  magnitude. Error bars are associated to standard errors.
              }         
         \label{CMb204}
  \end{figure}

In Fig. \ref{CMb204}  we plot (J-Ks) vs Ks colour--magnitude diagrams of the G21 galaxies and extended sources in b204 tile. We consider extinction corrected aperture magnitudes. In this plot we show with crosses G21 galaxies lost either by the (J-Ks) cut, the requirement of available photometry in all VVV bands or the limit in $\rm R_{1/2}$ SE star--galaxy separation parameter. Most G21 galaxies that do not fulfill $\rm R_{1/2}>$1 arcsec are located in the fainter region of the diagram. G21 galaxies lost by the colour cut present (J-Ks) colour close to our demarcation line. In G21 we restrict our selection of galaxy candidates to sources with (J-Ks)$>$0.97 mag, therefore the discrepancies in the galaxy colours is a consequence of both, differences in the photometry and in the extinction maps used in this work compared to G21. Also, G21 galaxies with no available photometry in at least one of the ZYJHKs bands are located in the faint region of the sample with mean Ks=15.8 mag. 

In the right region of this figure we also plot a densogram based on the (J-Ks) colour distribution of the sources and as can be seen most galaxies are grouped well above the adopted cut (J-Ks)=0.95 mag . In Fig. \ref{CMb204} we also show the fraction of recovered G21 galaxies, i.e. the completeness, as a function of Ks extinction corrected aperture magnitude. From this figure it can bee seen that completeness rate is almost 100\% for bright galaxies and it drops to levels closer to 80\% only for Ks$>$15 mag. Nevertheless, this is the magnitude range where we add 88 galaxy candidates to b204 tile. The large error bars in the bright bins are statically associated to the small number of extended objects with Ks$<$14 mag in G21 sample.

\section{Galaxy candidates in VVV Bulge}
\label{catalogue}
The results from the Previous section highlight the efficiency in our selection criteria to identify extended sources associated with galaxies in the VVV Bulge area.
Nevertheless, the process of identifying galaxy candidates in this  challenging sky region entails the downloading and processing of 191879 images in five photometric bands (each one of about 165 Kb). To this end we use the \texttt{wget} option of {multiget\_image VSA} tool to download ZYJHKs images of these sources. Taking into account the \texttt{wget} restriction on the maximum number of images, we split the data into 90 files containing each one about 2000 targets and  download ZYJHKs images of these sources. We use {AstroPy} \citep{Astropy2022} in a {Python} program to access the header of each image and extract important information as zero point magnitudes and seeing to be entered in the SE parameter file. Therefore we run SE on each individual image considering its specific header information and obtain the star--galaxy separation parameters that allows the decontamination of the sample from galactic sources. In this way we retain 38718 sources to be visually inspected in the search of extended objects. After visual inspection we obtain a sample of 14480  extended objects (hereafter VVV Bulge galaxies) and 24238 sources associated with stars in our Galaxy.

\subsection{The role of stellar density and Galactic absorption}

 \begin{figure*}
      \centering
       \includegraphics[width=0.99\textwidth]{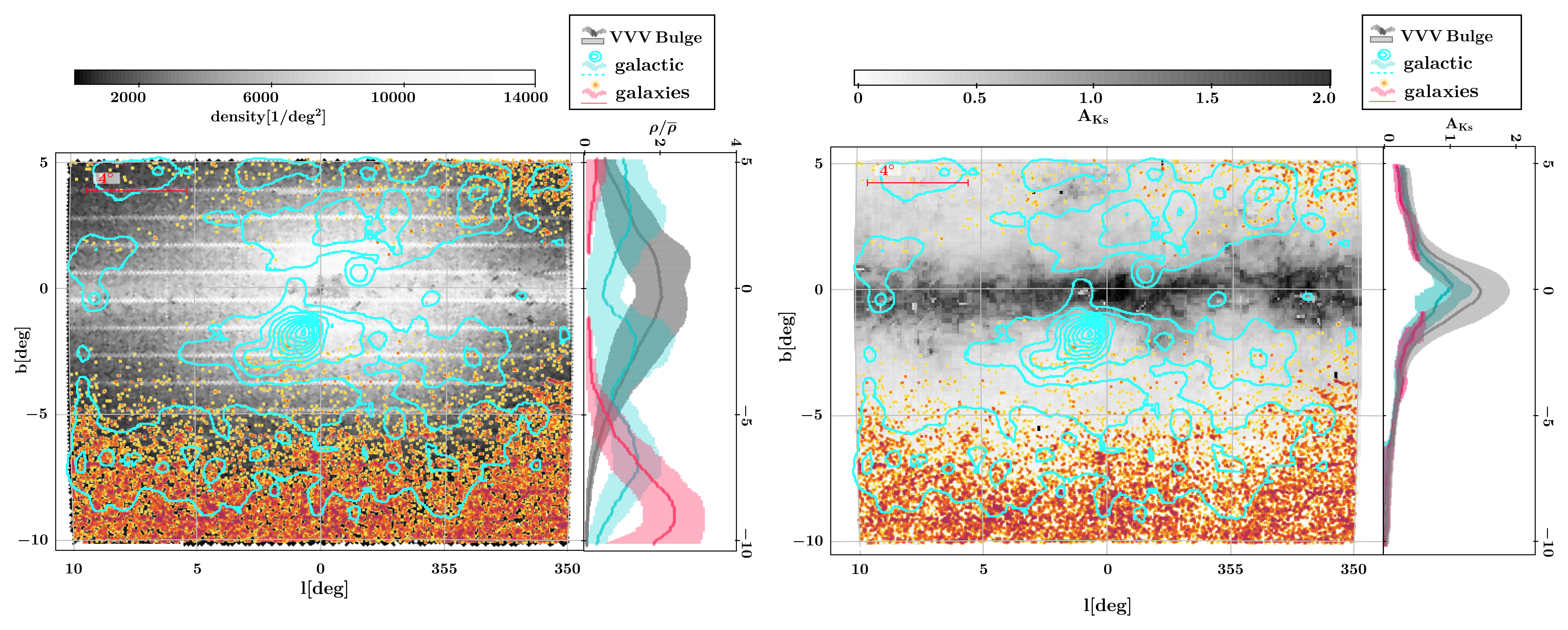}
      \caption{Left: Stellar density in VVV Bulge. We consider stars with Ks$\sim$13.0. Right:  Mean absorption in Ks band calculated considering a distance of 10 kpc.  The cyan contours represent 10 iso--density regions of Galactic sources in our targetsSE sample. The dots represents extended sources. 
              }
         \label{LB_stars_AKs}
  \end{figure*}

The VVV Bulge area is a complex region because the stellar density grows towards the galactic centre compromising the identification of extragalactic sources in crowded fields. Moreover the absorption levels at low galactic latitudes are as high as 2.5 magnitudes in Ks--band preventing the light of background galaxies to reaching us. 

We study the variation of the galactic sources (blended stars and saturated objects) and VVV Bulge galaxies in the targetsSE sample as a function of stellar-density and  Ks--band absorption.

To map the stellar density in the VVV Bulge, we select a sample of stars with Ks$\sim$13 mag. This magnitude was chosen in order to avoid bright objects and because for fainter sources magnitude the errors grow exponentially \citep{sai12}. In this way we select nearly 800000 stars in the  entire VVV-Bulge region.  To map galactic absorption we download Ks band absorption corrections from 3D map described in \cite{Chen2013}  considering a radius of 10 kpc to calculate the integrated absorption of the Milky Way. This map comprises 30200 pixels regularly distributed in the VVV-Bulge area.

In Fig. \ref{LB_stars_AKs} we show density distribution of the Ks$\sim$13 mag star sample. We pixelate the VVV-Bulge region in 7 acrmin x 7 arcmin pixels and calculate the star density in 1 square degree. The overlap regions between VVV tiles appear as regions of higher stellar density.
We also show the absorption map colour-codded according to Ks--band absorption values.   On these maps we over impose 10 iso--density contours representing the distribution of the galactic sources in the targetsSE sample. Also we plot the sky position of the extended objects that are galaxy candidates in the VVV Bulge.

This figure leads to a few conclusions. As expected, there is an increment of both absorption and stellar density toward low galactic latitude regions. 
We study the variation of the over--density ($\rho/\overline \rho$) of stars as a function of galactic latitude and compare this trend with that of galactic sources and VVV Bulge galaxies. We find that the density of stars increment towards low  galactic latitude but galactic sources in the targetsSE sample present a decrement in the distribution at $|\rm b|<$2 deg. In the case of galaxies there is a lack of these sources at that low latitudes and the density is highly increased for latitudes b$<$-5 deg (see right plot of left panel in Fig. \ref{LB_stars_AKs}).

We also study the variation of Ks mean galactic absorption ($A_{Ks}$) as a function of galactic latitude. We find an increment in the mean $A_{Ks}$ values toward low latitudes and a lack of extended objects associated with the absorption peak $A_{Ks}=2$ mag at $|\rm b|<$2 deg. On the other side, galactic sources follow the distribution of the \cite{Chen2013}  extinction map (see right plot of Fig. \ref{LB_stars_AKs} left panel).

As a conclusion of this section we find that, as expected, stellar crowding and Galactic absorption are strongly affecting the identification of galaxies behind the galactic Bulge.  Under our methodology, only when the stellar density is lower than the mean and the absorption  $A_{Ks}$ is lower than 0.5 magnitudes it is possible to identify galaxies in the VVV Bulge region with confidence. 

 \begin{figure*}
      \centering
       \includegraphics[width=0.75\textwidth]{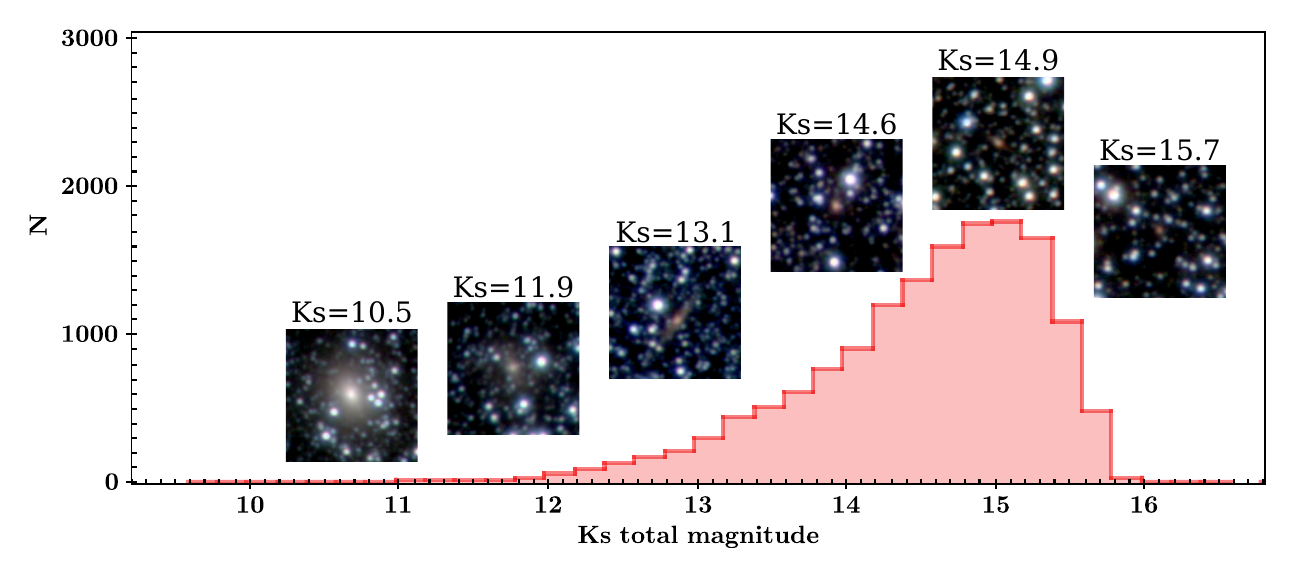}
      \caption{ Ks total magnitude distribution of VVV Bulge galaxies. We show 30 arcsec x 30 arcsec  KsJZ RGB stamp images of illustrative examples of galaxies in different magnitude bins.
              }
         \label{extended_examples}
  \end{figure*}

\subsection{Visual Inspection methodology}

The visual inspection  is a crucial step in the identification of galaxy candidates in VVV Bulge data. Therefore, we follow a specific and controlled procedure in this classification step. First we build KsJZ RGB images with WCS information and link these files to each target in our targetsSE catalogue. Then we use TOPCAT \citep{Topcat} to load our catalogue and display the images by using Activation Actions. We also verify the sources independently by Displaying HiPS cutout from VVV DR4 colour JYZ images, we perform this step to be sure that we are visually inspecting the correct object. The combination of JYZ images are not as suitable as KsJZ to identify extended sources but let us implement this independent verification step.  

The visual inspection was carried by 7 authors that were trained  to classify targetsSE as extended objects or galactic sources associated with  blended/saturated stars. To study the error in the classification all the astronomers classify a sample of 2000 random sources from the targetsSE catalogue. Based on this study we calculate a 7\% average error  associated with the visual inspection. This percentage is lower for brighter galaxies (4\%) and increases toward fainter objects reaching values closer to 10\%. Therefore we mostly expect  discrepancies for the faintest objects in our catalogue. In Fig. \ref{extended_examples} we show the total Ks  magnitude distribution of these sample together with illustrative examples of extended sources in different magnitudes bins. This figure shows how ambiguous may the visual inspection become for sources fainter than total Ks$\sim$15 mag.

As result of this study we find more than 14000 VVV Bulge galaxies that were identified as extended sources in the VVV Bulge region. Moreover we obtain SE photometry in ZYJHKs bands for all these sources, making this sample the largest galaxy catalogue to date in this area of the sky.

\subsection{A new picture of the Universe hidden behind the Galactic Bulge}

\begin{figure*}[]
    \centering
       \includegraphics[width=0.9\textwidth]{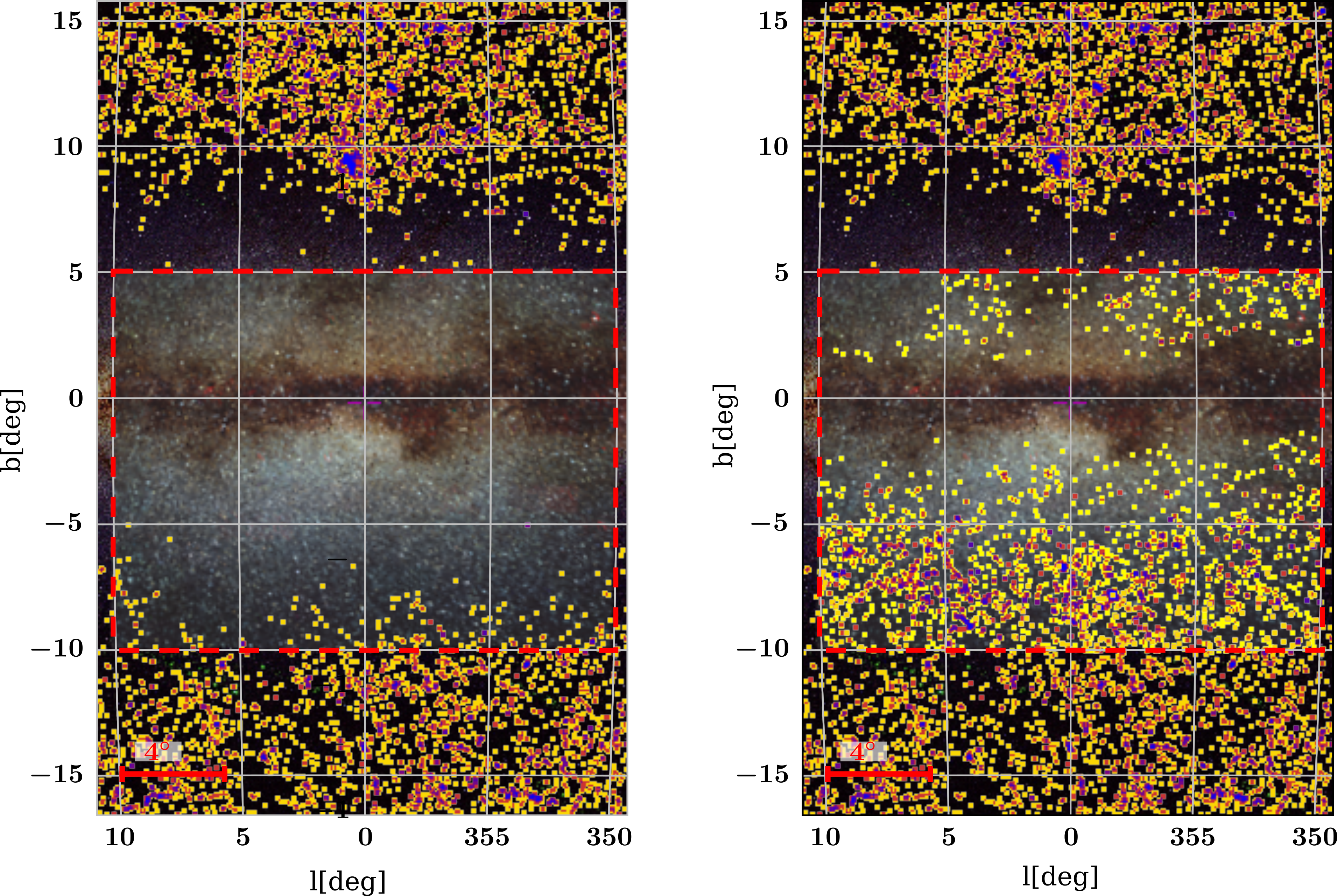}
    \caption{Distribution of galaxies  in a region of $|\rm l|<$10 deg $|\rm b|<$15 deg.  We show a RGB JYZ VVV image over--imposed on the false colour JHK 2MASS image. The left panel shows the distribution of 2MXSC galaxies and in the right panel we plot around 2000 VVV Bulge galaxy candidates with SE total magnitude Ks$<$13.5 mag.}
    \label{pictureUniverse}
\end{figure*}

The previous largest catalogue of galaxies in the VVV Bulge region was the 2MXSC. As stated in section \ref{c2m} there are only 271 galaxy--type 2MASS extended sources in the area of the sky under study in this paper. Moreover most 2MXSC galaxies are located at $|\rm b|>$7 deg. Therefore it is interesting to explore how the VVV Bulge galaxies help to complete the picture of Universe  at low galactic latitudes. To this end we study the distribution of galaxies  in a region of $|\rm l|<$10 deg $|\rm b|<$15 deg. We consider the magnitude limit of 2MASS sources and select galaxies brighter than Ks=13.5 mag, considering total magnitudes, from 2MXSC and also from our VVV Bulge galaxies sample.  In Fig. \ref{pictureUniverse} we plot a RGB JYZ VVV image over--imposed on the JHK 2MASS image. The left panel of the image shows the distribution of 2MXS galaxies and in the right panel we also plot around 2000 VVV Bulge galaxy candidates with SE total magnitude Ks$<$13.5 mag. From this figure it can be clearly appreciated how the catalogue of galaxies obtained in this work help to complete the picture on the Universe in this challenging area of the sky. The northern region at b$>$5 deg has been mapped by the Bulge-high VVVX survey and is currently under study.

\section{Summary and conclusions}
\label{conc}
In this work we define selection criteria to target extended sources in the VVV Bulge region ($|l|<10.0, -10.0<b<5.0$). We use the VSA facilities to extract relevant photometric information of catalogued sources and also to download images upon which we run SE to obtain specific star--galaxy separation parameters as well as suitable photometry of  extended sources. 

We find that initial photometric restrictions on sources from VSA \texttt{vvvSource} table combined with restrictions on star--galaxy separation parameters obtained from SE, is a successful strategy to achieve acceptable levels of contamination and a high completeness  in the construction of a sample suitable for the identification of galaxy candidates in the VVV Bulge. Moreover, based on the study of a representative random sample of 2000 objects obtained from the \texttt{vvvSource} table, we find that considering non-aperture corrected magnitudes APERMAG3, the restrictions:
   \begin{itemize} 
    \setlength\itemsep{0.2em}
      \item \textit{mergedClass=1}
      \item 10$<$Ks$<$16.2
      \item APERMAG3>0 in ZYJHKs bands
      \item J-H$>$0, H-Ks$>$0, J-Ks>0.95
      \item CS<0.5 and $\rm R_{1/2}>$1 arcsec
 \end{itemize}

 are the most suitable constraints to lower the initial contamination levels from 99\% to 60\% while retaining a completeness of about 80\% in the selection of extended sources. 

Under these restrictions we select 38718 targets from the $\sim$ 3.7 billion catalogued sources (10 mag$<$Ks$<$17 mag) in the \texttt{vvvSource} table. Nevertheless it is expected that only one third of these targets are galaxy--like.  Therefore,  visual inspection is crucial in the identification of galaxy candidates in VVV Bulge data and we follow a specific and controlled procedure in this classification step. We train 7 astronomers to classify targets into extended galaxy--like objects or blend/saturated sources associated with stars in our Galaxy. The visual inspection was carried on KsJZ false colour RGB images linked to each catalogued target. We use this specific bands combination in order to highlight red colour of galaxies against the bluer foreground stars. The visual inspection gave rise to a catalogue of more than 14000 galaxy candidates with total Ks magnitudes in the range 9.5 mag$<$Ks$<$16.5 mag. These sources are distributed avoiding the most crowded regions of the VVV Bulge and also considering that absorption hampers the identification at $|\rm b|<$2 deg. 

We test our target selection criteria with previous catalogued galaxies from the 2MXSC and G21 samples. We find that under our methodology we are able to recover 75\% of the catalogued 2MASS galaxies in the VVV Bulge region and 77\% of G21 galaxies in b204 tile. Therefore the completeness of our catalogue based in previous catalogued galaxies is higher than 75\%.

The catalogue presented in this work comprises 14480 galaxy candidates in the VVV Bulge region and it is the largest catalogue to date in the ZOA. Moreover as the VISTA photometry is compatible with 2MASS, our galaxy catalogue fills the region of the galactic Bulge ($|\rm l|<$10 deg) from b=-10 deg to b=-2 deg providing a homogeneous coverage of the region when considering galaxies with Ks$<$13.5 mag. Furthermore we are currently processing VVVX data to expand the coverage to higher Galactic latitudes.

Even if spectroscopic measurements are essential to map the position of these galaxy candidates, the results obtained in the present work provide a new picture of the Universe hidden behind the curtain of stars, dust and gas in the unexplored Milky Way Bulge region. 
  
Several comprehensive studies of this catalogue will be presented in  forthcoming  papers, including a detailed description of the properties of  galaxy candidates, the density environment at different scales from pairs to clusters, active galactic nuclei activity and also photometric redshifts estimation of a suitable sample of galaxies in our catalogue. A forthcoming paper will describe and present the full catalogue of VVV Bulge galaxies.

This paper, previous and future works demonstrate the potentiality of the VVV/VVVX survey to find and study a large number of galaxies and extragalactic structures obscured by the Milky Way, enlightening our knowledge of the Universe in this challenging and impressive region of the sky.

\begin{acknowledgements}
    FD specially thanks to Mike Read and VSA team for their helpful support.
    This work was supported in part by the Consejo Nacional de Investigaciones Cient\'ificas y T\'ecnicas de la Rep\'ublica Argentina (CONICET) and Secretar\'ia de Ciencia y T\'ecnica de la Universidad Nacional de San Juan.  The authors gratefully acknowledge data from the ESO Public Survey program ID 179.B-2002 taken with the VISTA telescope, and products from VISTA Science Archive (VSA). V.M. also acknowledges  support  from  project  DIDULS Regular N \textdegree PR2353857.
    This research made use of \texttt{Astropy},\footnote{http://www.astropy.org} a community-developed core \texttt{Python} package for Astronomy \citep{astropy:2013, astropy:2018} and \texttt{TOPCAT}- Tool for OPerations on Catalogues And Tables \citep{Topcat}.
    
\end{acknowledgements}

\section*{DATA AVAILABILITY}
The data underlying this article will be shared on reasonable request to the corresponding author.

\bibliographystyle{aa.bst}
\bibliography{bib}{}

\end{document}